\documentclass[prd,superscriptaddress,twocolumn,showpacs]{revtex4}
\usepackage{mathrsfs}
\usepackage{graphicx}
\usepackage{amssymb}
\newcommand{\D}{{\mathrm{D}}}

\newcommand{\be}{\begin{equation}}
\newcommand{\ee}{\end{equation}} 
\newcommand{\bea}{\begin{eqnarray}}
\newcommand{\eea}{\end{eqnarray}}
\newcommand{\ba}{\begin{array}}
\newcommand{\ea}{\end{array}}
\newcommand{\bra}[1]{\left(#1\right)}
\newcommand{\bras}[1]{\left[#1\right]}

\newcommand{\curl}{{\mathsf{curl}\,}}

\newcommand{\reff}[1]{(\ref{#1})}

\begin{document}

\title{Response to ``Comment on `Primordial magnetic seed field amplification by gravitational waves'\,"}

\author{Gerold Betschart}
\affiliation{Racah Institute of
Physics, Hebrew University of Jerusalem, Givat Ram, 91904 Jerusalem,
Israel}

\author{Caroline Zunckel}
\affiliation{Astrophysics, University of
Oxford, Denys Wilkinson Building, Keble Road, Oxford OX1 3RH, UK}

\author{Peter K S Dunsby}
\affiliation{Department of Mathematics
and Applied Mathematics,
    University of Cape Town, 7701 Rondebosch, South Africa}
\affiliation{South African Astronomical Observatory, Observatory
7925, Cape Town, South Africa}

\author{Mattias Marklund}
\altaffiliation[Also at: ]{Centre for Fundamental Physics, Rutherford Appleton Laboratory,
  Chilton, Didcot, Oxon OX11 OQX, U.K.}
\affiliation{Department of
Physics, Ume{\aa} University, SE-901 87 Ume{\aa}, Sweden}

\begin{abstract}
Here we respond to the comment by Tsagas \cite{T} on our paper
\cite{us}. We show that the results in that comment are flawed and
cannot be used for drawing conclusion about the nature of magnetic
field amplification by gravitational waves, and  give further
support that the results of \cite{us} are correct.
\end{abstract}

\maketitle
\section{Introduction}
In Ref.\ \cite{us} it was shown, using proper second order covariant
gauge-invariant perturbation theory, that the results in Ref.\
\cite{GWamp} concerning gravitational wave amplification of weak
magnetic fields gave amplification rates incorrect by several orders
of magnitude. The author of the comment \cite{T} claims that our
main result in Ref.\ \cite{us} does not apply outside the Hubble
radius, and that our numerical estimates and conclusions are
therefore invalid as a result of neglecting oscillatory terms.
Further claims that our approach is not truly gauge-invariant and
also not mathematically complete as a result of neglecting the issue
of constraints are also made. However, all these claims will be
refuted in what follows. Moreover, the comment contains misleading
statements (addressing some of our assumptions and approximations)
which are in contrast to what was done (and indeed clearly stated)
in our paper. In particular, we did not employ ideal
magnetohydrodynamics as the comment states; it was only assumed that
the electric field at first order is perturbatively smaller than the
magnetic field. That the gravito-magnetic interaction did not yield
an electric field at second order is a consequence of requiring an
homogeneous magnetic field interacting with gravity waves. Ideal
magnetohydrodynamics was only used in Ref.\ \cite{mhd}.

\section{The issue of the electric curl}
In section II of the comment \cite{T}, the author argues that
because of  Eq. (1) it is not possible {\it a priori} to assume that
$(\curl E_a)^{\dot{}}_{\perp}=0$ initially even though $\curl E_a
=0$ initially. However, it follows from Maxwell's equations that
 $(\curl E_a){\dot{}}_{\perp}=-\Theta \curl E_a + \curl \curl B_a$,
and since initially $\curl B_a$ vanishes (and therefore also $\curl
\curl B_a$) because the first-order magnetic field $\tilde B_a$ is
homogeneous, one concludes that also $(\curl
E_a)^{\dot{}}_{\perp}=0$ when the interaction between gravitational
waves and the first-order magnetic field is turned on. When these
constraints are used in the governing equation for $\curl E_a$ (see
Eq. 2 in \cite{T}), one indeed finds that in the spatially flat case
the generated electric field stays curl-free if it was initially so.

The author of the comment came to the opposite conclusion by
using the relation
\be \label{relation}
\curl \curl B_a = -\D^2 B_a + {\cal R}_{ab}B^b .
\ee
However, $B_a$ is not a gauge-invariant variable at second
order, and thus cannot be used for the purpose of analyzing the
appropriateness of the initial conditions. Let us make this important
point very clear: the usual splitting of the
magnetic field $B_a$ into a first-order homogeneous part and a
second-order term, $B_a=\tilde B_a + B_a^{(2)}$, as the author of
the comment assumed, leads to critical inconsistencies when commutation relations
are involved (see also Eqs. (15) and (16) in \cite{us}). Taking the
spacetime background to be spatially flat and inserting the said
splitting in the right hand side of Eq. (\ref{relation}) yields
\be
 \curl \curl B_a = -\D^2 B_a^{(2)} + {\cal R}_{ab}^{(1)}\tilde
B^b,\label{r1}
\ee
where ${\cal R}_{ab}^{(1)}$ denotes the first-order contribution to
the 3-Ricci tensor. On the other hand, if one first employs the splitting
for the term $\curl B_a$, and then applies Eq. \reff{relation},
one obtains
\bea
 && \curl \curl B_a = \curl \curl B_a^{(2)} \nonumber \\ && \qquad
  = -\D^2 B_a^{(2)} + {\cal R}_{ab}^{(0)}B^b_{(2)} 
  = -\D^2 B_a^{(2)},\label{r2}
\eea
which differs from the result \reff{r1}. This ambiguity, due to the gauge problem with the
magnetic field splitting in relation to the use of the commutator relation, renders Eq.
(1) in \cite{T} meaningless and invalidates the conclusion drawn
from it.

We stress that we neglected the electric current  at all orders in
our paper \cite{us}, which is \textit{potentially} physically unsound.
However, the inclusion of a first-order current requires the
inclusion of a first-order inhomogeneous magnetic field, which was
beyond the scope of the paper \cite{us}. Currents and velocity
perturbations have been explicitly taken into account in our
follow-up paper \cite{mhd} by making use of ideal
magnetohydrodynamics. In this case we did indeed have $(\curl E^a) \neq
0$, it was shown that this term only contributes at very small
scales to the generated magnetic field. On large scales, the results
for the generated magnetic field agree  with the ones found
previously (compare Eq. (50) of \cite{mhd} with Eq. (49) of
\cite{us}).

\section{Scales and amplification}
When discussing super-horizon scales in section III of his comment,
the author correctly points out that the solution Eq. (7) in Ref.
[2] of the interaction term $I(\tau)$ in the case of dust is valid
as long as $x \equiv 2 \,\ell\, \tau^{1/3}/(a_0H_0)\ll 1$ holds
(similarly for the case of radiation). However, since the
gravitational wave number is defined as $\ell =
2\,\pi\,a/\lambda_{GW}$ and the Hubble length $\lambda_H=1/H$, we
also have $x=2\,\pi(\lambda_H/\lambda_{GW})_0\,\tau^{1/3}$, and
therefore the condition $x\ll 1$ will sooner or later break down as
$\tau$ grows, even though the ratio $(\lambda_H/\lambda_{GW})_0$ is
small for a particular gravity wave mode with wavenumber $\ell$. The
solutions Eq. (4) and Eq. (5) in \cite{T} therefore describe the
evolution of an initially super-horizon magnetic mode [meaning
$(\lambda_H/\lambda_{GW})_0\ll 1$] only correctly up to the time of
horizon crossing (at best). Thereafter one has to rely on the
solutions Eqs. (49) and (50) given in \cite{us}.

\begin{figure}
  \includegraphics[width=7cm]{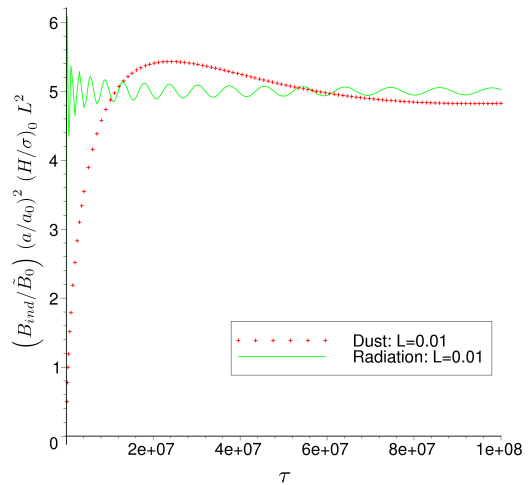}\\
  \caption{The dimensionless induced magnetic fields plotted against dimensionless
  time $\tau$ for dust and radiation, taking $L =0.01$.
  For given initial conditions the generated magnetic field saturates
  at a
  value described by our result \reff{summary}. The super-adiabatic
  growth is approximately depicted by the infinite-wavelength solutions
  (46) and (47) in \cite{us}.}  \label{plot1}
\end{figure}

The fact that we did not display the solutions of the generated
magnetic field for the dust and radiation case in full detail when
dealing with the  modes with non-zero wavenumber, $\ell \neq 0$,
seems to have led to some confusion. To arrive at the solutions Eqs.
(49) and (50) in \cite{us} we used generic initial conditions
leading to complicated expressions, and they were thus not stated
explicitly in \cite{us}; however, all the relevant terms for the
amplification have been given and the nature of the non-displayed
expressions has been circumscribed in \cite{us}. To improve on this,
we give in the following the full solution for the \emph{total}
(first- plus second-order) magnetic field in the radiation era
assuming initial conditions
$I_{(\ell)}(\tau=1)=\sigma^{(\ell)}_0\tilde B_0$ and
$I'_{(\ell)}(1)=0$ for the interaction variable. The latter
condition simply means that $(\dot
\sigma_{(\ell)}/H)_0=2\sigma_0^{(\ell)}$ has been chosen. With these
initial conditions, Eq. (50) in \cite{us} gives
\bea
  && B^{(\ell)}_{\mathrm{Rad}}(a) = \tilde
  B_0\,\bra{\frac{a_0}{a}}^2\Bigg\{ 1+
  \bra{\frac{\sigma}{H}}_0\frac{5}{L^2}
\nonumber \\ && \,\,
  + \frac{a_0}{a}\bra{\frac{\sigma}{H}}_0
  \Bigg[ \cos\bra{L\frac{a}{a_0}}\frac{5\sin L -L^2\sin L -5L\cos
  L}{L^3}
\nonumber \\ &&\qquad
  +\sin\bra{L\frac{a}{a_0}}\frac{L^2\cos L -5L\sin L -5\cos
  L}{L^3} \Bigg]\Bigg\}, \label{mfrad}
\eea
where $L \equiv \ell/(a_0H_0) = 2\pi(\lambda_H/\lambda_{GW})_0$. As
expected, the \emph{generated} magnetic field is proportional to the
initial shear anisotropy $(\sigma/H)_0$; it starts to grow from a
zero value until it \emph{saturates} to a constant value (modulo
adiabatic decay); this is also evident from FIG. \ref{plot1}. Note
that the solution \reff{mfrad} is exact and valid for all finite
wavelengths at all times. In the infinite-wavelength limit, $\ell
\rightarrow 0$, the solution for the magnetic field goes over into
\bea
  &&\frac{B^{(0)}_{\mathrm{Rad}}(a)}{\tilde
  B_0} = \bra{\frac{a_0}{a}}^2\Bigg\{1+\frac23\bra{\frac{\sigma}{H}}_0
  \bras{\frac{a_0}{a}-1}
\nonumber \\   && \qquad
  +\frac56\bra{\frac{\sigma}{H}}_0\bras{\bra{\frac{a}{a_0}}^2-1} \Bigg\},
\label{longrad}
\eea
which is of course the same as Eq. (47) in \cite{us}, where
identical initial conditions had been used.

A numerical analysis of the gravito-magnetic interaction (cf. FIG.
\ref{plot1}) demonstrates that the generated magnetic field starts
from zero, undergoes super-adiabatic growth until it begins to
oscillate around a value which depends on the initial conditions but
is the same during both the radiation or dust/reheating eras. Whilst
the super-adiabatic growth is approximately described by the
infinite-wavelength solutions Eqs. (46) and (47) in \cite{us}, the
saturation effect is only present in the finite-wavelength solutions
Eqs. (49) and (50) in \cite{us} [see also Eq. \reff{mfrad} above].

What then is the emerging magnetic field at the end of the day? Returning
to Eq. \reff{mfrad} and dividing by the energy density of the
background radiation, one obtains (the oscillatory parts decay away)
\be
\frac{B}{\mu^{1/2}_{\gamma}} \simeq
 \bras{1+\frac{1}{10}
\bra{\frac{\lambda_{\tilde B}}{\lambda_{\mathrm{H}}}}^2_0
\bra{\frac{\sigma}{H}}_0}\bra{\frac{\tilde
B}{\mu^{1/2}_{\gamma}}}_0\ , \label{summary}
\ee
where the wavenumber indices have been suppressed and the resonant
condition $\lambda_{\mathrm{GW}}\sim \lambda_{\tilde B}$ --- the
gravity wavelength $\lambda_{\mathrm{GW}}$ matches the size $
\lambda_{\tilde B}$ of the magnetic region --- has been used. This
agrees with our result Eq. (51) given in \cite{us}. As pointed out
in our paper, the case of dust is very similar and leads to the same
result \reff{summary} for the maximally resulting magnetic field.

It is plain to see that in the light of the above (being in unison with
comments already made in our paper \cite{us}), the criticism brought
forward by the author of the comment (see Eqs.\ (8) and (9) therein)
is not correct. Also, Eq.\ (9) of \cite{T} is
algebraically wrong and can not be used to draw conclusions
about the validity of the results in \cite{us}.

In the last paragraph of section III of the comment, the author
claims that our numerical results have been compromised because we
did not take reheating effects into account. He further claims that
one cannot use the result Eq. \reff{summary} for this purpose and
presents instead formula Eq. (10) in \cite{T}. However, that formula
is of third order, and is therefore not correct since the
gravito-magnetic interaction is of second order. If one assumes that
reheating does not last long enough in order for the induced
magnetic field to reach saturation, one may use Eq. (5) of \cite{T}
to find the field at the end of reheating, which may then be
employed in our result Eq. \reff{summary} to yield the total
magnetic field during the radiation era. Including as always only
terms up to second order, we obtain
\be
\frac{B(a)}{\tilde B_0} \simeq \,\bra{\frac{a_0}{a}}^2\!\!\bras{1 + 2\!
\bra{\frac{\sigma}{H}}_0\!\! \frac{a_{\mathrm{RH}}}{a_0} +\frac{1}{10}\!
\bra{\frac{\lambda_{\mathrm{\tilde
B}}}{\lambda_{\mathrm{H}}}}^2_{\mathrm{RH}}\!\!\!
\bra{\frac{\sigma}{H}}_{\mathrm{RH}} } , \label{corr}
\ee
correcting Eq. (10) of \cite{T}.

Contrary to the claim of the comment,  one can very well use our result
\reff{summary}, which was shown in \cite{us} to hold for the
radiation as well as the dust/reheating era, in order to take
reheating into account. Let us for simplicity assume that reheating
lasts long enough such that the generated magnetic field saturates
during the reheating phase. In a first step, we employ the result
\reff{summary} to find the total field at the end of reheating,
\be
\frac{B_{\mathrm{RH}}}{\tilde
B_0}  \simeq \bra{\frac{a_0}{a_{\mathrm{RH}}}}^2\bras{1+\frac{1}{10}
\bra{\frac{\lambda_{\mathrm{\tilde B}}}{\lambda_{\mathrm{H}}}}^2_0
\bra{\frac{\sigma}{H}}_0} . \label{reh}
\ee
Here, the suffix $0$ denotes the time at the end of inflation when
reheating starts. In a second step we use again Eq. \reff{summary}
but this time with $B_{\mathrm{RH}}$ as input magnetic field. We
then find that the total magnetic field during the radiation era is
given by
\be
\frac{B(a)}{\tilde B_0} \simeq \bra{\frac{a_0}{a}}^2\!\!\bras{1+\frac{1}{10}\!
\bra{\frac{\lambda_{\mathrm{\tilde B}}}{\lambda_{\mathrm{H}}}}^2_0\!\!\!
\bra{\frac{\sigma}{H}}_0\! +\frac{1}{10}\!
\bra{\frac{\lambda_{\mathrm{\tilde
B}}}{\lambda_{\mathrm{H}}}}^2_{\mathrm{RH}}\!\!\!
\bra{\frac{\sigma}{H}}_{\mathrm{RH}} } , \label{rehrad}
\ee
where a third-order term has been discarded. In our article
\cite{us} the second gravitational wave term in square brackets in
\reff{rehrad} was carefully calculated and found to be of order
$\sim 10^{-6}$, as expected from a perturbation calculation
\footnote{Observe that in section V. of \cite{us} the suffix $0$
denoted the end of reheating, whereas here it has the suffix $RH$.}.
The first gravitational wave term in the square bracket represents
the reheating correction and since it is of the same perturbative order as
the term previously discussed, its value will not be much different
(in any case $\leq 1$). The same comment applies when one uses Eq.
\reff{corr} instead of Eq. \reff{rehrad}. The claim that our
numerical results have been compromised is thus refuted.

\section{Gauge-invariance and linearity}
The author of the comment claims that obtaining the second-order generated magnetic
field via integration of the gauge-invariant variable
$\beta_{a}=\dot B_{<a>} + 2\Theta B_a/3$ is in itself not a
gauge-invariant procedure. However, if we expand the magnetic field
$B_a = B_a^{(1)} +B_a^{(2)} +\ldots$ into first, second and higher
order parts, then we obtain $\beta_a=\dot B_{<a>}^{(2)} + 2\Theta
B_a^{(2)}/3$, and $\beta_a$ thus evidently only describes the
second-order part (higher order terms are neglected as usual). To
get $B_a^{(2)}$ from an integration is harmless since, as we pointed
out in section  II above and also in section IIC of our paper
\cite{us}, the gauge issue arises if commutator relations have to be
calculated, which is not the case here.

\begin{figure}[t]
  \includegraphics[width=7cm]{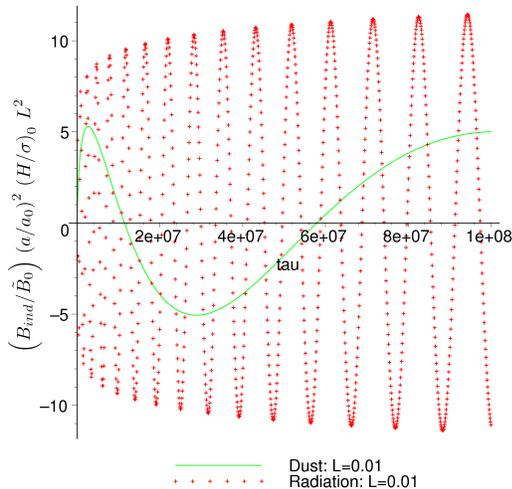}\\
  \caption{The generated magnetic field in the weak-field approximation in dimensionless
  units. The induced magnetic fields are plotted against the dimensionless
  time variable $\tau$ for the dust and radiation cases
  taking $L \equiv \ell/(a_0H_0) = 2\pi(\lambda_H/\lambda_{GW})_0=0.01$.  }\label{plot2}
\end{figure}

The author of the comment further writes in the second paragraph of
section IV \cite{T} that there was no new information in our
second-order approach in comparison to his own first-order approach
to the problem at hand \cite{GWamp}. This makes little sense, since
the two methods are fundamentally different in nature, and also
since the governing equations for the generated magnetic field
differ in both cases. Moreover, when solving numerically the
relevant weak-field equations [see Eqs. (61) and (62) in \cite{us}]
for the magnetic field, taking identical initial conditions as
discussed above, one obtains different results (cf. FIG.
\ref{plot2}). In contrast to the magnetic field obtained in our
second-order approach (see FIG. \ref{plot1}), the weak-field
magnetic field shows large oscillations but no saturation.

\section{Summary}
In this response to the criticism brought forward by the comment
\cite{T} of Tsagas, we pointed out the following:

  $\bullet$ The author's argumentation regarding the curl of the electric field
  is untenable since it is based on a relation (Eq. (1) in \cite{T}) which is not
  valid at second order;

  $\bullet$ The setting we employed was \emph{not} that of ideal
  magnetohydrodynamics;

  $\bullet$ Further strengthening our main results in \cite{us} with numerical experiments
  demonstrates that there was no improper assessment of
  scale in \cite{us};

  $\bullet$ Equation (9) in \cite{T} is algebraically wrong, and can thus not be used for
  drawing conclusions about the validity of the expressions in \cite{us};

  $\bullet$ Although our formalism explicitly makes use of second order perturbation theory,
  the author of the comment claims that our numerical results
  are comprised because we neglected a third-order term [see
  (10) in \cite{T}];

  $\bullet$ The generated magnetic field obtained by means of the
  weak-field approximation (see \cite{T}) differs significantly from the one
  obtained with our formalism, the former shown to give rise to unreasonable
  amplification rates;

  $\bullet$ The second-order variable $\beta_{a}=\dot B_{<a>} + 2\Theta
  B_a/3$ contains only the second-order magnetic field $B_a^{(2)}$,
  and that an integration of $\beta_a$ with respect to time yields
  the evolution of $B_a^{(2)}$.

In conclusion, the criticism contained in the comment has thus been
shown to be unfounded. Rather, it should be pointed out that, if
treated using the proper gauge-invariant covariant second order
perturbation theory, the interaction between magnetic fields and
gravitational waves can give rise to interesting effects and be a
possible source of a boost of weak seed fields as energy is
transfered between the different degrees of freedom.



\end{document}